\documentclass[pra,superscriptaddress,nofootinbib,twocolumn,showpacs,preprintnumbers,amsmath,amssymb]{revtex4}

\usepackage{graphicx}
\usepackage{subfigure}

\newcommand{\ie}{i.\,e.\ }
\newcommand{\Ie}{I.\,e.\ }
\newcommand{\eg}{e.\,g.\ }

\newtheorem{assumption}{Assumption}
\newtheorem{result}{Result}

\begin{document}

\title{Minimal assumption derivation of a Bell-type inequality}

\author{Gerd Gra{\ss}hoff} \email{gerd.grasshoff@philo.unibe.ch}
\affiliation{History and Philosophy of Science, Exact Sciences, Sidlerstrasse,
   University of Bern, CH-3012 Bern, Switzerland}
\homepage{http://www.philoscience.unibe.ch}
\author{Samuel Portmann}
\email{portmann@itp.unibe.ch} \affiliation{History and Philosophy of Science,
Exact Sciences, Sidlerstrasse,
   University of Bern, CH-3012 Bern, Switzerland}
\affiliation{Institute of
   Theoretical Physics, University of Bern, Switzerland}
\homepage{http://www-itp.unibe.ch}
\author{Adrian W\"uthrich}
\email{awuethr@itp.unibe.ch} \affiliation{History and Philosophy of Science,
Exact Sciences, Sidlerstrasse,
   University of Bern, CH-3012 Bern, Switzerland}
\affiliation{Institute of
   Theoretical Physics, University of Bern, Switzerland}

\date{\today}

\begin{abstract}
   John Bell showed that a big class of local hidden-variable models 
 stands in 
   conflict with quantum mechanics and experiment.  Recently, there were
   suggestions that empirical adequate hidden-variable models might exist, 
   which presuppose a weaker notion of local causality. We will show 
   that a Bell-type inequality can be derived also from these weaker
   assumptions.
\end{abstract}


\maketitle

\section{Introduction}

The violation of Bell's inequality by the outcome of an EPR-type spin 
experiment
\cite{epr,bohm51} seems to exclude a local theory with hidden variables. The
underlying \emph{reductio ad absurdum} proof infers on the grounds of the
empirical falsification of the derived inequality that at least one of the required
assumptions must be false. The force of the argument requires that the
derivation be deductive and that all assumptions be explicit. We aim to 
extract 
a minimal set of assumptions needed for a deductive derivation of Bell's
inequalities given perfect correlation of outcomes of an EPR-type spin
experiment with parallel settings.

One of the assumptions in Bell's original derivation \cite{bell64} was
determinism. Later, he succeeded in deriving a similar inequality without
determinism \cite{bell71}, placing in its stead an assumption later 
dubbed \emph{local causality} \cite{bell76}. As Bell stressed, the 
notion of local causality he and others used might be challenged. In
\cite{redei99} it was pointed out, that Reichenbach's Common Cause Principle
\cite{reichenbach56} indeed suggests a weaker form of local causality. We will
prove here, however, that even from this weaker notion Bell's inequality can
still be derived.\footnote{Several of the issues we present in this paper are
   discussed in more detail in \cite{wuethrich04}.}

\section{The EPR-Bohm experiment}\label{epr}

\begin{figure}
   \includegraphics[width=\linewidth]{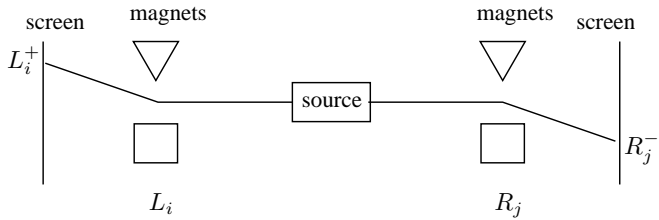}
\caption{\label{fig:eprb}Setup of the EPR-Bohm experiment. Cf.\
   \cite[p.~140]{bell87}.}
\end{figure}
Consider the so-called \emph{EPR-Bohm (EPRB) experiment} 
\cite{epr,bohm51}. Two
spin-$\frac{1}{2}$ particles in the \emph{singlet state}
\begin{align}
   |\Psi\rangle =\frac{1}{\sqrt{2}}\left( |\!\uparrow\downarrow\rangle
     -|\!\downarrow\uparrow\rangle\right)
\end{align}
are separated in such a way that one particle moves to a measurement apparatus
in the left wing of the experimental setting and the other particle to a
measurement apparatus in the right wing (see FIG.~\ref{fig:eprb}). The
experimenter can choose arbitrarily one of three directions in which the 
spin is
measured with a Stern-Gerlach magnet.

The following terminology follows the reconstruction by 
\textcite{wigner70} and which \textcite{vanfraassen89} has
subsequently expanded on. The event 
type\footnote{We
will speak of \emph{event types} to distinguish them from the
   \emph{token events} which instantiate corresponding event types.} that the
left (right) measurement apparatus is set to measure the spin in
direction~$i\in\{1,2,3\}$ is symbolized by $L_i$ ($R_i$).  $L_i^a$ ($R_i^a$)
symbolizes the event type that the measurement outcome in the left (right) 
wing
of a spin measurement in direction $i$ is $a$. There are two possible
measurement outcomes \emph{spin up} ($a\!=\!+$) and \emph{spin down} 
($a\!=\!-$)
for each particle in each direction. The letter $j\in\{1,2,3\}$ will be used
like $i$ to symbolize directions and $b\in\{+,-\}$ like $a$ to symbolize
measurement outcomes.  Formulas in which the variables $i$, $j$, $a$, and $b$
appear are meant to hold---if not otherwise stated---for all possible 
values of
the variables.  $p(X)$ denotes the probability of an event type $X$, which is
empirically measurable as the relative frequency of all runs of an EPRB
experiment in which the event type $X$ is instantiated, with respect to all
runs.  $p(X\wedge Y)$ is the probability of the event type `$X$ and $Y$',
measurable as the relative frequency of all runs in which both $X$ and $Y$ are
instantiated.  $p(X|Y)=p(X\wedge Y)/p(Y)$ is the conditional probability 
of the
event type $X$ \emph{given} the event type $Y$, measurable as the relative
frequency of instantiations of $X$ with respect to the subensemble of all runs
in which $Y$ is instantiated. E.g.
\begin{align}
   p(L_i^a\wedge R_j^b| L_i\wedge R_j)
\end{align}
denotes the probability that the measurement outcome is $a$ on the left 
and $b$
on the right, when measuring in direction $i$ on the left and in direction $j$
on the right. These probabilities are predicted by quantum mechanics as 
\begin{align}
   p(L_i^+\wedge R_j^+| L_i\wedge
   R_j)&=\frac{1}{2}\sin^{2}\frac{\varphi_{ij}}{2}, \label{cor1}\\
   p(L_i^-\wedge R_j^-| L_i\wedge
   R_j)&=\frac{1}{2}\sin^{2}\frac{\varphi_{ij}}{2}, \label{cor2}\\
   p(L_i^+\wedge R_j^-| L_i\wedge
   R_j)&=\frac{1}{2}\cos^{2}\frac{\varphi_{ij}}{2}, \label{cor3}\\
   p(L_i^-\wedge R_j^+| L_i\wedge
   R_j)&=\frac{1}{2}\cos^{2}\frac{\varphi_{ij}}{2}, \label{cor4}
\end{align}
where $\varphi_{ij}$ denotes the angle between the two measurement directions
$i$ and $j$. Also, the outcomes on each side are predicted separately to be 
completely random:
\begin{align}
   p(L_i^a |L_{i}\wedge R_{j})=\frac{1}{2}, \label{rand1}\\
   p(R_j^b |L_{i}\wedge R_{j})=\frac{1}{2}. \label{rand2}
\end{align}

\section{Local causality}

The derivations of Bell-type inequalities known to us which \emph{do not
   presuppose determinism} assume instead what John Bell calls \emph{local
   causality} \cite{bell76,horne74}. That is, the assumption that there is 
common cause variable\footnote{For the sake of simplicity, we assume that 
this partition 
   is discrete and finite. As will become clear in the following, the
   derivation of Bell's inequality can also be done without this restriction.}
$V$ which takes on values $q\in I=\{q_1,q_2,q_3,\dots,q_k\}$ such that 
for event 
types `the variable $V$ has the value $q$' ($V\!q$) we have $\sum_q p(V\!q)=1$
and
\begin{align}\label{eq:loc-caus}
   p(L_i^a\wedge R_j^b|V\!q\wedge L_i\wedge R_j) &= p(L_i^a|V\!q\wedge
   L_i)\nonumber \\
   &\quad \times p(R_j^b|V\!q\wedge R_j).
\end{align}
Other frequently used names for this condition are \emph{factorizability}
\cite{butterfield89} and \emph{strong locality} \cite{jarrett84,jarrett89}. 
It is
usually justified by pointing out that it follows from the conjunction of the
following three conditions, which are called \emph{completeness}
(equation~(\ref{scr-off})) and \emph{locality} (equations~(\ref{hloc1})
and~(\ref{hloc2})) \cite{jarrett84,jarrett89}, \emph{outcome independence} and
\emph{parameter independence} \cite{shimony93}, or \emph{causality} and
\emph{hidden locality} \cite{vanfraassen89}:
\begin{align}
   p(L_i^a\wedge R_j^b|V\!q\wedge L_i\wedge R_j)
   &=p(L_i^a|V\!q\wedge L_i\wedge R_j)\nonumber\\
   &\quad \times p(R_j^b|V\!q\wedge L_i\wedge
   R_j),\label{scr-off}\\
   p(L_i^a|L_i\wedge R_j\wedge V\!q)&=
   p(L_i^a|L_i\wedge V\!q), \label{hloc1} \\
   p(R_j^a|L_i\wedge R_j\wedge V\!q)&= p(R_j^a|R_j\wedge V\!q). \label{hloc2}
\end{align}
Equation~(\ref{scr-off}) says that event types $V\!q$ or the variable $V$
\emph{``screens off''} $L_i^a$ and $R_j^b$ from each other
\cite{vanfraassen89,butterfield89}. Van Fraassen \cite{vanfraassen89} pointed
out, that equation~(\ref{scr-off}) can be motivated through \emph{Reichenbach's 
   Principle of Common Cause} (PCC) \cite{reichenbach56}. The principle
states, that whenever two different event types $A$ and $B$ are statistically
correlated
\begin{align}
   p(A\wedge B)\neq p(A) p(B)
\end{align}
and neither $A$ is causally relevant for $B$ nor $B$ for $A$, there exists a
common cause variable $V$ with values $q\in I=\{q_1,q_2,q_3,\dots,q_k\}$
($\sum_q p(V\!q)=1$) such that $A$ and $B$ given $V\!q$ are uncorrelated:
\begin{align}
   p(A\wedge B|V\! q)= p(A|V\! q) p(B|V\! q).
\end{align}
In its original formulation the principle is stated only for a common cause
event type $C$, which is included in our formulation as the special case where
$V\! q$ can take only two values: $V\! q_{1}=C$, $V\! q_{2}=\neg C$ (`not 
$C$').
The principle was formulated for general common cause variables by
\textcite{redei03} and \textcite{placek00b}. Besides the screening-off 
condition
\textcite{reichenbach56} and \textcite{redei03} stipulate further restrictions
on the common cause variable, which are, however, irrelevant for our purposes.
\\
Now, as can be seen from equations~(\ref{cor1})-(\ref{cor4}), the event type
$L_{i}^{a}$ is in general correlated with event type $R_{j}^{b}$. It is
\begin{align}
   p(L_{i}^{a}|L_{i}\wedge R_{j})&=p(R_{j}^{b}|L_{i}\wedge R_{j})=\frac{1}{2},
\end{align}
and therefore
\begin{align}
   p(L_{i}^{a}\wedge R_{j}^{b}|L_{i}\wedge R_{j})&\neq p(L_{i}^{a}|L_{i}\wedge
   R_{j}) p(R_{j}^{b}|L_{i}\wedge R_{j}) \nonumber \\
   \mbox{except for}\qquad \varphi_{ij}&=\frac{\pi}{2} \mod \pi .
\end{align}
Supposing that $L_{i}^{a}$ is not causally relevant for $R_{j}^{b}$ and vice
versa (which is reinforced by the fact that the setup of the experiment can be
chosen so that the instantiations of $L_{i}^{a}$ and $R_{j}^{b}$ in each 
run 
of the experiment are space-like separated), PCC requires a common cause 
variable
which fulfills equation~(\ref{scr-off}). There are several different 
correlations;
e.g.~$L_{1}^{+}$ is correlated with $R_{2}^{+}$, and $L_{2}^{+}$ is correlated
with $R_{3}^{+}$. For each of these correlations PCC enforces the 
consequence 
that a common cause variable exists. As stressed in \cite{redei99} nothing
in PCC dictates that the common cause variables of the different 
correlations 
have to be the same. However, in all the derivations of Bell's inequality 
known
to us this identification is made nevertheless. It is further shown in
\cite{redei99} and \cite{redei03}, that for any set of correlations it is
mathematically possible to construct common cause variables. The authors
concluded in \cite{redei99} that the apparent contradiction between this
possibility and the claim that the EPRB correlations do not allow for a common
cause variable \cite{vanfraassen89,butterfield89}, is resolved by pointing out
that in the derivation of Bell's inequality a \emph{common} common cause
variable for all measurements is assumed:
\begin{quote}
   ``The crucial assumption in the [...] derivation of the [Clauser-Horne]
   inequality is that [the two-valued common cause variable] is a [two-valued
   common cause variable] \emph{for all four} correlated pairs, \ie that
   [$V\!q$] is a \emph{common} common cause [variable], shared by different
   correlations.  Without this assumption Bell's inequality \emph{cannot} be
   derived. But there does not seem to be any obvious reason why common causes
   should also be common common causes, whether of quantum or of any other 
 sort
   of correlations.'' (Italics in the original.)
\end{quote}

Showing the mathematical possibility of constructing common cause 
variables for any set of correlations and in particular for the correlations
found in the EPRB experiment is not sufficient for proving the existence 
of a physically
``natural'' hidden-variable model for that experiment, however. Besides being common 
cause variables (thus fulfilling equation (\ref{scr-off})), parameter
independence should hold, too (equations~(\ref{hloc1}) and~(\ref{hloc2})). Also, they 
should not be correlated with the measurement choices. As shown by Szab\'o
\cite{Szabo:1998pb}, it is possible to construct a model which fulfills these
requirements for each of the common cause variables separately. However, the
\emph{conjunctions} and other logical combinations of the event types that 
the 
common cause variables have certain values correlate in that model with the
measurement operations.  Whether a model can be constructed without these
correlations was posed as an open question by Szab\'o. This question is answered 
negatively by
the derivation of Bell's inequality that we present in the remainder of this
article.

\section{Bell's inequality from separate common causes}

\subsection{A weak screening-off principle}

Consider an EPRB experiment where the same direction $i$ ($i\in 
\{1,2,3\}$) is chosen in both wings. That is, in each run the event type 
$L_i\wedge R_i$ is 
instantiated. With this special setting quantum mechanics predicts (see 
equations~(\ref{cor1})-(\ref{rand2}), with $\varphi_{ij}=0$) that the measurement
outcomes in each wing are random but that the outcomes in one wing are 
perfectly
correlated with the outcomes in the other wing: if and only if the spin of 
the 
left particle is up, then the spin of the right particle is down, and vice 
versa. We
refer to this assumption as \emph{perfect correlation}, or PCORR for short.
\begin{assumption}[PCORR]
\begin{equation}
\label{perf-corr5}
   p_{ii}(R_i^-|L_i^+) = 1\ \mbox{and}\ \
   p_{ii}(L_i^+|R_i^-) = 1.
\end{equation}
\end{assumption}
We use here the definition
\begin{align}
   p_{ij}(\dots )\doteq p(\dots |L_{i}\wedge R_{j}).
\end{align}
Large spatial separation of coinciding events of type $L_i^a$ and $R_j^b$
suggests that the respective instances are indeed distinct events. This 
excludes
an explanation of the correlations by \emph{event identity}, as is the 
case, for 
example, with a tossed coin for the perfect correlation of the event types 
`heads up' and
`tails down'. Such a perfect correlation is explained in 
that every instance of `heads up' is also an instance of `tails
down', and vice versa. Since the separation is even space-like, no $L_i^a$ 
or $R_j^b$ should be causally relevant for the other. We refer to these two 
assumptions as \emph{separability}, SEP for short, and \emph{locality~1} 
(LOC1).

\begin{assumption}[SEP]\label{sep5}
   The coinciding instances of $L_i^a$ and $R_j^b$ are distinct events.
\end{assumption}

\begin{assumption}[LOC1]\label{loc1-5}
   No $L_i^a$ or $R_j^b$ is causally relevant for the other. 
\end{assumption}

Rather, there should be a common cause variable; that is, we assume PCC.
\begin{assumption}[PCC]\label{pcc-hv}
   If two event types $A$ and $B$ are correlated and the correlation cannot be
   explained by direct causation nor event identity, then there exists a 
 common
   cause variable $V$, with values $q\in I=\{q_1,q_2,q_3,\dots,q_k\}$ such 
 that
   $\sum_q p(V\!q)=1$ and
   \[
   p(A\wedge B|V\!q)=p(A|V\!q)p(B|V\!q),\qquad \forall q.
   \]
\end{assumption}
As already mentioned, we omit the other Reichenbachian conditions
\cite{reichenbach56,redei03} since they are not necessary for our derivation.

This principle together with the assumptions~PCORR, SEP and LOC1 implies that
there is for each of the EPRB correlations a (separate) common cause variable
$V^{+-}_{ij}$ with $q\in I^{+-}_{ij}$.
\begin{result}
  \begin{align}
\label{sep-scr-off}
   p_{ii}(L_i^+\wedge R_i^-|V^{+-}_{ii}q) &=
   p_{ii}(L_i^+|V^{+-}_{ii}q) \nonumber\\
   &\quad\times p_{ii}(R_i^-|V^{+-}_{ii}q).
  \end{align}
\end{result}
Note that common cause variables can be different for different correlations.

\subsection{Perfect correlation and ``determinism''}\label{sec:perf-corr-det}

We now show that from the fact that a \emph{perfect} correlation is 
screened off
by some variable it follows that without loss of generality the common cause
variable can be assumed to be two-valued and that the having of one of 
the two
values of the variables is necessary and sufficient for the instantiation 
of the
two perfectly correlated event types, cf.~\cite{suppes76}.

Let $A$ and $B$ be perfectly correlated,
\begin{displaymath}
   p(A|B)=p(B|A)=1,
\end{displaymath}
and screened-off from each other by a common cause variable,
\begin{displaymath}
   p(A\wedge B|V\!q)=p(A|V\!q) p(B|V\!q).
\end{displaymath}
We can split the set $I$ of all values $V$ completely into two disjunct 
subsets, namely in the subset $I^+$ of those values of $V$ for which 
$p(A\wedge
V\!q)$ is not zero and in the subset $I^-$ of those for which it is zero:
\begin{gather*}
   I^+ = \{q\in I:\ p(A\wedge V\!q)\neq 0\}, \\
   I^- = \{q\in I:\ p(A\wedge V\!q) =0\}, \\
   I=I^-\cup I^+,\ I^-\cap I^+ = \emptyset.
\end{gather*}
 >From this definition of $I^-$ it follows already that
\begin{equation}
   \label{eq:i+-nec-a}
   p(A|V\!q)=0,\quad \forall q\in I^-,
\end{equation}
\ie that $V\!q$ with $q\in I^+$ is necessary for $A$. Moreover, for all $q\in
I^+$ we have by screening off and perfect correlation 
\begin{equation}
   \label{eq:i+-suff-a}
   p(A|V\!q)=p(A|B\wedge V\!q)=1.
\end{equation}
That the variable has a value in $I^+$ is a necessary and sufficient condition
for $A$. The following calculation shows that $V\!q$ with $q\in I^+$ is also
necessary and sufficient for $B$.

 >From perfect correlation it follows that
\begin{displaymath}
   p(B|A\wedge V\!q)=1,\quad \forall q\in I^+.
\end{displaymath}
That $V\!q$ screens off $B$ from $A$ yields
\begin{displaymath}
   p(B|A\wedge V\!q)=p(B|V\!q).
\end{displaymath}
Together with the previous equation this implies that $V\!q$ is sufficient for
$B$ for all $q\in I^+$:
\begin{equation}
   \label{eq:i+-suff-b}
   p(B|V\!q)=1\quad \forall q\in I^+.
\end{equation}

If $q\in I^-$ we have by definition $p(A\wedge V\!q)=0$, which implies
\begin{displaymath}
   p(A\wedge B\wedge V\!q)=0.
\end{displaymath}
By perfect correlation we have therefore also $p(B\wedge V\!q)=0$, which 
in turn
implies that
\begin{equation}
   \label{eq:i+-nec-b}
   p(B|V\!q)=0,\quad \forall q\in I^-,
\end{equation}
which means that $V\!q$ with $q\in I^+$ is also necessary for $B$.

This calculation shows that in the case of a perfect correlation the set of
values of the common cause variable decomposes into two relevant sets. This
means that whenever there is an (arbitrarily-valued) common cause variable 
for a 
perfect correlation, there is also a two-valued common cause variable, 
namely the
disjunction of all event types $V\!q$ for which $q\in I^+$ or $q\in I^-$,
respectively.
\begin{align*}
   C & = \vee_{q\in I^+} V\!q, \\
   \neg C & = \vee_{q\in I^-} V\!q.
\end{align*}
We refer to $C$ as a \emph{common cause event type}. In the case of a perfect
correlation no generality\label{no-generality} is achieved by allowing for a
more than two-valued common cause variable; if there is a common cause 
variable
for a perfect correlation, there is also a common cause event type. Moreover,
the common cause event type is a necessary and sufficient condition for the
event types that are screened off by it (equations~\eqref{eq:i+-nec-a},
\eqref{eq:i+-suff-a}, \eqref{eq:i+-suff-b} and \eqref{eq:i+-nec-b}).

Result 1 thus implies that there is a common cause event type 
$C_{ii}^{+-}$ such
that
\begin{gather}
   p_{ii}(L_i^+|C_{ii}^{+-}) = p_{ii}(R_i^-|C_{ii}^{+-}) = 1,
   \label{cii-nec-suff-1}\\
   p_{ii}(L_i^+|\neg C_{ii}^{+-}) = p_{ii}(R_i^-|\neg C_{ii}^{+-}) = 0.
   \label{cii-nec-suff-2}
\end{gather}
The sub- and superscripts of $C_{ii}^{+-}$ refer to $C_{ii}^{+-}$ being the
common cause event type of $L_i^+$ and $R_i^-$.

The outcome of a spin measurement is always either $+$ or $-$ and nothing 
else. We call this assumption \emph{exactly one of exactly two possible
   outcomes} (EX).
\begin{assumption}[EX]\label{ass:ex}
   \begin{alignat}{2}
     p_{ii}(L_i^+) + p_{ii}(L_i^-) &= 1, & \qquad p_{ii}(L_i^+\wedge L_i^-) &=
     0,
     \label{ex-1-otc-1}\\
     p_{ii}(R_i^+) + p_{ii}(R_i^-) &= 1, & \qquad p_{ii}(R_i^+\wedge R_i^-) &=
     0.
     \label{ex-1-otc-2}
  \end{alignat}
\end{assumption}
As stressed by \textcite{fine82}, among the actual measurements there are 
always 
runs in which no outcome is registered, which is normally attributed to the
limited efficiency of the detectors and not taken to the statistics. If one
assumes instead, that part of these no-outcome runs are caused by the hidden
variable, then it is possible to construct empirically adequate models for the
EPRB experiments \cite{szabo00, szabo02}. With assumption~\ref{ass:ex}, we
explicitly exclude such models.

With assumption~\ref{ass:ex}, while $C_{ii}^{+-}$ is necessary and 
sufficient for $L_i^+$ and $R_i^-$, its complement, \ie $\neg C_{ii}^{+-}$ is
necessary and sufficient for the opposite outcomes, \ie $L_i^-$ and $R_i^+$:
\begin{gather}
   p_{ii}(L_i^-|C_{ii}^{+-}) = p_{ii}(R_i^+|C_{ii}^{+-}) = 0,
   \label{opp-otc-1}\\
   p_{ii}(L_i^-|\neg C_{ii}^{+-}) = p_{ii}(R_i^+|\neg C_{ii}^{+-}) = 1.
   \label{opp-otc-2}
\end{gather}

\subsection{A \emph{minimal theory} for spins}\label{sec:min-th}

In section~\ref{sec:perf-corr-det} it was found that $C_{ii}^{+-}$ is 
sufficient
for $L_i^+$ \emph{given} parallel settings ($L_i\wedge R_i$), see
equation~\eqref{cii-nec-suff-1}. \Ie the conjunction $C_{ii}^{+-}\wedge
L_i\wedge R_i$ is sufficient for $L_i^+$. But because of space-like separation
of events of type $L_i^+$ and $R_i$ that are instantiated in the same run, the
latter types should not be causally relevant for the former. The measurement
choice in one wing should be causally irrelevant for the outcomes (and the
choices) in the other wing. Therefore we should discard $R_i$ from the
sufficient conjunction. The part $C_{ii}^{+-}\wedge L_i$ \emph{alone} is
sufficient for $L_i^+$. A similar reasoning can be applied to $R_j^+$, 
$R_j$ and
$\neg C_{jj}^{+-}$, cf.\ equation~\eqref{opp-otc-2}. This is our assumption
\emph{locality~2} (LOC2).
\begin{assumption}[LOC2]\label{loc2-5}
   If $L_i\wedge R_i\wedge X$ is sufficient for $L_i^+$, then $L_i\wedge X$
   alone is sufficient for $L_i^+$; and similarly for $R_j^+$, \ie if
   $L_j\wedge R_j\wedge Y$ is sufficient for $R_j^+$, then $R_j\wedge Y$ alone
   is sufficient for $R_j^+$.
\end{assumption}
Moreover, the remaining part $C_{ii}^{+-}\wedge L_i$ is \emph{minimally}
sufficient, in the sense that none of its parts is sufficient on its
own.\footnote{Minimal sufficient conditions as definied by \cite{grasshoff01}
   and \cite{gg_baumgartner}.} If, for example, $C_{11}^{+-}$ is instantiated,
but we do not choose to measure $L_1$, then $L_1^+$ will not be instantiated.
That is to say, we cannot discard yet another conjunct of $L_i\wedge
C_{ii}^{+-}$ as we discarded $R_i$ from $C_{ii}^{+-}\wedge L_i\wedge R_i$.

Let us turn to \emph{necessary} conditions for $L_i^+$. To begin with, 
$L_i$ is
necessary: If there is no Stern-Gerlach magnet properly set up ($L_i$) the 
particle is not deflected either up- or downwards; similarly for $L_i^-$, 
$R_j^+$
and $R_j^-$. Roughly speaking, \emph{no outcome without measurement} (NOWM).
\begin{assumption}[NOWM]
   \begin{alignat}{2}
     p(L_i^+\wedge\neg L_i) &= 0, & \qquad p(L_i^-\wedge\neg L_i) &= 0,
     \label{eq:no-outc-wo-meas-1} \\
     p(R_j^+\wedge\neg R_j) &= 0, & \qquad p(R_j^-\wedge\neg R_j) &= 0.
     \label{eq:no-outc-wo-meas-2}
   \end{alignat}
\end{assumption}

Second, we saw in section~\ref{sec:perf-corr-det}
that if parallel settings are chosen and $\neg C_{ii}^{+-}$ is
instantiated an event of type $L_i^+$ does never occur. In other words, $\neg
C_{ii}^{+-}\wedge L_i\wedge R_i$ implies $\neg L_i^+$:
\begin{equation}\label{eq:loc3-new}
 \neg C_{ii}^{+-}\wedge L_i\wedge R_i \rightarrow \neg L_i^+.
\end{equation}\index{$\rightarrow$ (implies)}
Again we propose a locality condition based on the idea that the measurement
choice in one wing should be causally irrelevant\index{causally irrelevant}
for the outcomes (and the choices) in the other wing:\footnote{The following
 version of LOC3 is slightly different from an earlier version of the
 article. We thank Gabor Hofer-Szab\'o, Miklos R\'edei and I\~{n}aki San Pedro for
 their comments.} If $\neg C_{ii}^{+-}\wedge L_i\wedge R_i$
 are sufficient for $\neg L_i^+$, then $\neg C_{ii}^{+-}\wedge L_i$ \emph{alone}
should be sufficient for $\neg L_i^+$. A similar reasoning can be applied to
$R_j^+$, $R_j$ and $C_{jj}^{+-}$, cf.\ equation~\eqref{opp-otc-1}. 

\begin{figure}
     \subfigure[]{\label{fig:mth-a}%
       \includegraphics[width=.35\linewidth]{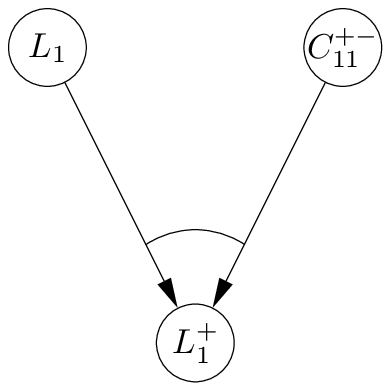}} \hspace{1cm}
\subfigure[]{\label{fig:mth-b}%
   \includegraphics[width=.35\linewidth]{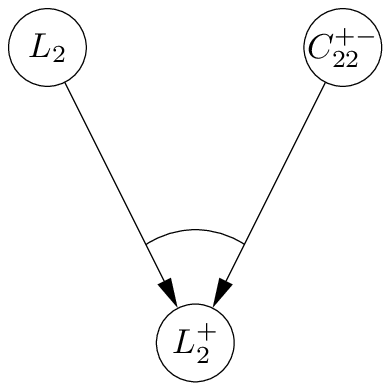}}
\subfigure[]{\label{fig:mth-c}%
   \includegraphics[width=.35\linewidth]{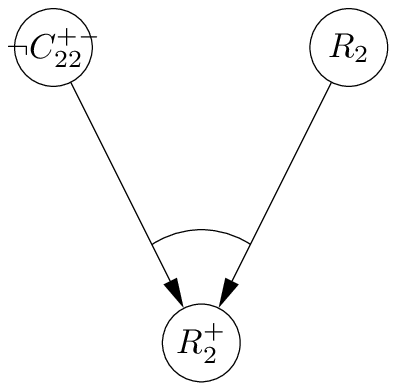}} \hspace{1cm}
\subfigure[]{\label{fig:mth-d}%
   \includegraphics[width=.35\linewidth]{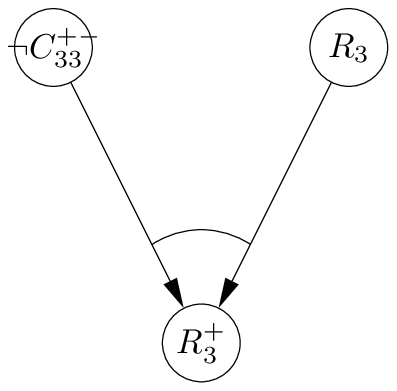}}
     \caption{Minimal theories for outcomes of spin measurements.}
     \label{fig:min_th}
\end{figure}

\begin{assumption}[LOC3]\label{loc3-5}
If $L_i\wedge R_i\wedge X$ is sufficient for $\neg L_i^+$,
 then $L_i\wedge X$ alone is sufficient for $\neg L_i^+$; and
 similarly for $\neg R_j^+$, \ie if $L_j\wedge R_j\wedge Y$ is sufficient
 for $\neg R_j^+$, then $R_j\wedge Y$ alone is sufficient for
 $\neg R_j^+$.    
\end{assumption}
By LOC3 it follows from equation~\eqref{eq:loc3-new} that
\begin{equation}
  \label{eq:loc3-new-appl}
    \neg C_{ii}^{+-}\wedge L_i \rightarrow \neg L_i^+.
\end{equation}
This is equivalent to
\begin{equation}
  \label{eq:loc3-new-appl-equiv}
    L_i^+\wedge L_i \rightarrow C_{ii}^{+-},
\end{equation}
and also to
\begin{equation}
  \label{eq:loc3-new-appl-equiv-2}
    L_i^+\wedge L_i \rightarrow C_{ii}^{+-}\wedge L_i.
\end{equation}
According to equation~\eqref{eq:no-outc-wo-meas-1}, $L_i$ is necessary for
$L_i^+$. That means $L_i^+\rightarrow L_i$, but also $L_i^+\rightarrow
L_i^+\wedge L_i$. Above, we have found (eq.~\eqref{eq:loc3-new-appl-equiv-2})
that $L_i^+\wedge L_i \rightarrow C_{ii}^{+-}\wedge L_i$.  Altogether, this
entails $L_i^+\rightarrow L_i\wedge C_{ii}^{+-}$, \ie that $L_i\wedge
C_{ii}^{+-}$ is necessary for $L_i^+$.
 Moreover, it is a \emph{minimally} necessary condition in the 
sense of
\cite{grasshoff01} since it does not contain any disjuncts. All in all:
$C_{ii}^{+-}\wedge L_i$ is a minimally necessary and minimally sufficient
condition for $L_i^+$. In a similar vein we find that $R_j\wedge\neg
C_{jj}^{+-}$ is minimally necessary and minimally sufficient for $R_j^+$. 
We have 
thus derived in particular the four \emph{minimal theories} in the
sense of
\cite{grasshoff01} as illustrated in FIG.~\ref{fig:min_th}.

In a formal notation the four minimal theories read as the following four
equations, where $\leftrightarrow$ is the usual \emph{biconditional}, which
means that the left-hand side implies the right-hand side and vice
versa.\footnote{For details see \cite{grasshoff01} and \cite{gg_baumgartner}.
   Note in particular that a correct formal notation of a minimal theory uses
   what Gra{\ss}hoff~et.~al. \cite{grasshoff01,gg_baumgartner}\ call a
   \emph{double conditional}.} This intermediate result is referred to as
\emph{minimal theories} (MTH).
\begin{result}[MTH]\label{min-th}
\begin{align*}
   (L_1\wedge C_{11}^{+-}) &\leftrightarrow L_1^+, \tag{\ref{fig:mth-a}} \\
   (L_2\wedge C_{22}^{+-}) &\leftrightarrow L_2^+,\tag{\ref{fig:mth-b}} \\
   (R_2\wedge \neg C_{22}^{+-}) &\leftrightarrow R_2^+, 
 \tag{\ref{fig:mth-c}} \\
   (R_3\wedge \neg C_{33}^{+-}) &\leftrightarrow R_3^+. \tag{\ref{fig:mth-d}}
\end{align*}
\end{result}

From the logical relations (\ref{fig:mth-a}), (\ref{fig:mth-b}), (\ref{fig:mth-c})
and (\ref{fig:mth-d}) the following probabilities can be derived:
\begin{align*}
   p(L_1^+\wedge R_2^+) &= p(L_1\wedge C_{11}^{+-}\wedge R_2\wedge \neg
   C_{22}^{+-}), \\
   p(L_2^+\wedge R_3^+) &= p(L_2\wedge C_{22}^{+-}\wedge R_3\wedge \neg
   C_{33}^{+-}), \\
   p(L_1^+\wedge R_3^+) &= p(L_1\wedge C_{11}^{+-}\wedge R_3\wedge \neg
   C_{33}^{+-}).
\end{align*}
By NOWM (equations~\eqref{eq:no-outc-wo-meas-1} and
\eqref{eq:no-outc-wo-meas-2}) $p(L_1^+\wedge R_2^+)$ is the same as
$p(L_1^+\wedge R_2^+\wedge L_1\wedge R_2)$ etc. and the above equations read:
\begin{align}
   p(L_1^+\wedge R_2^+\wedge &L_1\wedge R_2)\nonumber\\
   &= p(L_1\wedge C_{11}^{+-}\wedge R_2\wedge \neg C_{22}^{+-}),
     \label{probs1}\\
     p(L_2^+\wedge R_3^+\wedge &L_2\wedge R_3)\nonumber\\
     &= p(L_2\wedge C_{22}^{+-}\wedge R_3\wedge \neg C_{33}^{+-}),
     \label{probs2}\\
     p(L_1^+\wedge R_3^+\wedge &L_1\wedge R_3)\nonumber\\
     &= p(L_1\wedge C_{11}^{+-}\wedge R_3\wedge \neg
     C_{33}^{+-}).\label{probs3}
\end{align}

\subsection{No conspiracy}\label{sec:no-cons-new}

The events of type $C_{ii}^{+-}$ are not supposed to be influenced by the
measuring operations $L_i$ and $R_j$. One reason for this assumption is 
that the
measurement operations can be chosen arbitrarily before the particles 
enter the
magnetic field of the Stern-Gerlach magnets and that an event of type
$C_{ii}^{+-}$ is assumed to happen \emph{before} the particles arrive at the
magnets. Therefore a causal influence of the measurement operations on 
events of
type $C_{ii}^{+-}$ would be tantamount to \emph{backward causation}. Also an
inverse statement is supposed to hold: The event types $C_{ii}^{+-}$ are 
assumed
not to be causally relevant for the measurement operations. This is meant to
rule out some kind of ``cosmic conspiracy'' that whenever an event of type
$C_{ii}^{+-}$ is instantiated, the experimenter would be ``forced'' to use 
certain 
measurement operations. This \emph{causal} independence between $C_{ii}^{+-}$
and the measurement operations is assumed to imply the corresponding
\emph{statistical} independence. The same is assumed to hold also for
\emph{conjunctions} of common cause event types. We refer to this condition as
\emph{no conspiracy} (NO-CONS).
\begin{assumption}[NO-CONS]
     \begin{equation}
\label{no-cons5}
     p(C_{ii}^{+-}\wedge \neg C_{jj}^{+-}|L_i\wedge R_j) = p(C_{ii}^{+-}\wedge
     \neg C_{jj}^{+-}).
   \end{equation}
\end{assumption}

By this condition of statistical independence the three probabilities 
considered
above can be transformed. That is, we have, for instance
   \begin{eqnarray*}
       p(L_1^+\wedge R_2^+|L_1\wedge R_2)
       &\doteq & \frac{p(L_1^+\wedge R_2^+\wedge L_1\wedge R_2)}{p(L_1\wedge
       R_2)} \\
     &\stackrel{\mbox{\small (i)}}{=}& \frac{p(L_1\wedge C_{11}^{+-}\wedge
       R_2\wedge \neg C_{22}^{+-})}{p(L_1\wedge R_2)} \\
       &\doteq & p(C_{11}^{+-}\wedge \neg C_{22}^{+-}|L_1\wedge R_2) \\
       &\stackrel{\mbox{\small (ii)}}{=}& p(C_{11}^{+-}\wedge \neg 
 C_{22}^{+-})
       \\
       &\stackrel{\mbox{\small (iii)}}{=}& p(C_{11}^{+-}\wedge \neg
       C_{22}^{+-}\wedge C_{33}^{+-}) \\
       & & + p(C_{11}^{+-}\wedge \neg C_{22}^{+-}\wedge \neg C_{33}^{+-}).
   \end{eqnarray*}
   The dotted equations are true by definition of conditional probability.  In
   step~(i) equation~\eqref{probs1} was used.  Step~(ii) is valid by ``no
   conspiracy'' (equation~\eqref{no-cons5}), and (iii) by a theorem of
   probability calculus, according to which $p(A)=p(A\wedge B)+p(A\wedge\neg
   B)$ for any $A$ and $B$.  Transforming the other two expressions in a
   similar way, we arrive at
\begin{align}
   p(L_1^+\wedge R_2^+&|L_1\wedge R_2) \nonumber\\
   &= p(C_{11}^{+-}\wedge \neg C_{22}^{+-}\wedge C_{33}^{+-})\nonumber\\
   &\qquad + p(C_{11}^{+-}\wedge \neg
   C_{22}^{+-}\wedge \neg C_{33}^{+-}), \label{cprobs1}\\
   p(L_2^+\wedge R_3^+&|L_2\wedge R_3) \nonumber\\
   &= p(C_{11}^{+-}\wedge C_{22}^{+-}\wedge \neg
   C_{33}^{+-})\nonumber\\
   &\qquad + p(\neg C_{11}^{+-}\wedge C_{22}^{+-}\wedge \neg C_{33}^{+-}),
   \label{cprobs2}
\end{align}
\hrulefill
\begin{align}
   p(L_1^+\wedge R_3^+&|L_1\wedge R_3) \nonumber\\
   &= p(C_{11}^{+-}\wedge C_{22}^{+-}\wedge \neg
   C_{33}^{+-})\nonumber\\
   &\qquad + p(C_{11}^{+-}\wedge \neg C_{22}^{+-}\wedge \neg C_{33}^{+-}).
   \label{cprobs3}
\end{align}
Since both terms of the right-hand side of the last equation appear in the sum
of the right-hand sides of the first two equations, the following version 
of the
Bell inequality (BELL) follows\footnote{It was first derived in this form
   by Wigner \cite{wigner70}.}.
\begin{result}[BELL]
\begin{align}
\label{bell5} p(L_1^+\wedge R_3^+|L_1\wedge R_3) &\leq p(L_1^+\wedge
R_2^+|L_1\wedge R_2)\nonumber\\ &\qquad + p(L_2^+\wedge R_3^+|L_2\wedge R_3).
\end{align}
\end{result}
This inequality has been empirically falsified, see e.\,g.~\cite{aspect82}.
\\
The inequality was derived from the following assumptions.
\begin{itemize}\label{list5}
\item Perfect correlation (PCORR),
\item separability (SEP),
\item locality~1 (LOC1),
\item principle of common cause (PCC), 
\item exactly one of exactly two possible outcomes (EX),
\item locality~2 (LOC2),
\item no outcome without measurement (NOWM),
\item locality~3 (LOC3),
\item no conspiracy (NO-CONS).
\end{itemize}
This is a version of Bell's theorem. It says: If these assumptions are 
true, the
Bell inequality is true. The derivation of the Bell inequality presented 
here is
an improvement on the usual Bell-type arguments, such as \cite{bell76} and
\cite{vanfraassen89}, in two respects: First, it does not assume a 
\emph{common}
common cause variable for different correlations. Second, contrary to the usual
locality conditions, the ones assumed here do not presuppose a solution to the
problems posed by the relation between causal and statistical (in)dependence
(see \eg~\cite{spirtes93}).

\section{Discussion}

Our claim to have presented a minimal assumption derivation of a Bell-type
inequality is relative: our set of assumptions is weaker than any set known to
us from which a Bell-type inequality can be derived and that contains the
assumption of \emph{perfect} correlation (PCORR). It was one of the 
achievements
of \textcite{horne74} to show that a Bell-type inequality can be derived 
also if
the correlations of outcomes of parallel spin measurements are not assumed to
be
perfect. Our
assumption of correlation is stronger than the one used by Clauser and Horne.
However, they assume a \emph{common} common cause variable for all 
correlations, 
which is a stronger assumption than our assumption of possibly different 
common
cause variables for each correlation (PCC). We have not been able to derive a
Bell-type inequality without assuming perfect correlation and allowing
different common cause variables. If PCORR is indeed a necessary
assumption for our derivation of the Bell inequality, it should be possible to
construct a model in which PCORR does not hold (being violated by an arbitrary
small deviation, say). Since the actually measured correlations are never
perfect---a fact that is usually attributed to experimental imperfections---it
is not obvious how such a model could be refuted.

Our notion of local causality might be challenged as follows. Even though
nothing in PCC dictates that in general the common cause variables of 
different correlations have to be the same, there might be strong grounds 
for why
they are the same in the context of the EPRB experiment. Indeed, Bell 
argued for
his choice of local causality along the following lines.\footnote{For a
   very good and more detailed discussion of this, see \cite{butterfield89}.}
Assume that $L_i^a$ and $R_j^b$ are positively correlated.  Then
\begin{align}
   p(L_{i}^{a}|R_{j}^{b}\wedge L_{i}\wedge R_{j})>p(L_{i}^{a}|L_{i}\wedge
   R_{j}).
\end{align}
Since coinciding instances of $L_i^a$ and $R_j^b$ are space-like separated,
neither is causally relevant for the other. Rather, the correlation should be
explained by exhibiting some common causes in the overlap of the backward 
light
cones of the coinciding instances. An instance of, say, $L_i^a$ raises the
probability of an instantiation of one of the common causally relevant 
factors,
and this raises the probability of an instantiation of $R_j^b$. But 
\emph{given
the total state} of the overlap of the backward light cones of two coinciding
instances, the probability of, say, $R_j^b$ is assumed to be the same whether 
$L_i^a$ is instantiated or not. If the total state of the overlap of the
backward light cones is already given, nothing more that could be causally
relevant for $R_j^b$ can be inferred from an instance of $L_i^a$.

\begin{figure}
   \includegraphics[width=.9\linewidth]{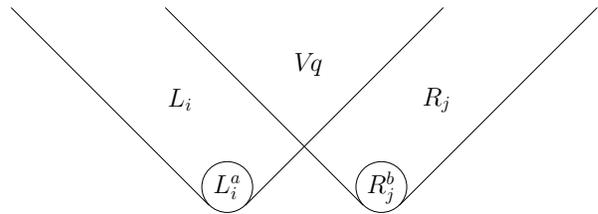}
\caption{\label{fig:cones}The two backward light cones of two measurement
   outcomes. The total state of the overlap is taken to define a common common
   cause variable $V$, which can take on certain values $q$. Cf.\
   \cite[p.~55]{bell87}}.
\end{figure}

Along this line of reasoning the total state $V$ of the overlap of the 
backward
lightcones\footnote{One might argue that the total state of the
   \emph{union} of the backward lightcones is a better candidate for a common
   cause variable \cite{butterfield89}. The following discussion carries 
 over also to
   this case.} of $L_i^a$ and $R_{j}^{b}$ is a common cause variable
which screens off the correlation:
\begin{align}
   p(L_{i}^{a}\wedge R_{j}^{b}|L_{i}\wedge R_{j}\wedge
   V\! q)&=p(L_{i}^{a}|L_{i}\wedge R_{j}\wedge V\! q) \nonumber \\
   &\quad \times p(R_{j}^{b}|L_{i}\wedge R_{j}\wedge V\! q).
\end{align}
The common past $V\!q$ cannot be altered by choosing one or the other 
direction
for the spin measurement---\emph{``facta infecta fieri non
   possunt''} \cite[p.~185]{placek00b}. Therefore the total state
$V\!q$ of the common past is indeed a \emph{common} common cause variable for
all correlated outcomes, see FIG.~\ref{fig:cones}.

This reasoning can be questioned along the following lines. It is 
reasonable that 
not all event types that are instantiated in the overlap of the backward light
cones of two coinciding instances of the correlated event types are 
causally 
relevant for these latter event types. Therefore conditionalizing on the 
total state is
conditionalizing not only on the relevant factors but also on the irrelevant.
Moreover, it is conceivable that which event types of the common past are
relevant and which are not differs for different measurements. Claiming 
that the
total state of the common past is a common common cause variable, one is thus
committed to assume that
\begin{quote}
   ``conditionalizing on all other events [\dots] in addition to those
   affecting [the correlated event types], does not disrupt the stochastic
   independence induced by conditionalizing on the affecting events.''
   \cite{butterfield89}
\end{quote}
In particular in the light of \emph{Simpson's paradox} \cite{simpson51} this
assumption has been challenged \cite{cartwright79}. Here, we will not assess
arguments in favour of or against the possibility that conditionalizing on 
irrelevancies yields unexpected statistical dependencies. Our point is 
that by 
weakening the assumption in the way we did, our derivation is conclusive 
whatever
may be the answer to this question.

\begin{acknowledgments}
   We would like to thank Guido Bacciagaluppi, Mikl\'os R\'edei, Gabor Hofer-Szab\'o and I\~{n}aki San Pedro  for 
 fruitful
   discussions.

\end{acknowledgments}

\newpage 

\bibliography{phys-rev}

\begin{thebibliography}{30}
\expandafter\ifx\csname natexlab\endcsname\relax\def\natexlab#1{#1}\fi
\expandafter\ifx\csname bibnamefont\endcsname\relax
  \def\bibnamefont#1{#1}\fi
\expandafter\ifx\csname bibfnamefont\endcsname\relax
  \def\bibfnamefont#1{#1}\fi
\expandafter\ifx\csname citenamefont\endcsname\relax
  \def\citenamefont#1{#1}\fi
\expandafter\ifx\csname url\endcsname\relax
  \def\url#1{\texttt{#1}}\fi
\expandafter\ifx\csname urlprefix\endcsname\relax\def\urlprefix{URL }\fi
\providecommand{\bibinfo}[2]{#2}
\providecommand{\eprint}[2][]{\url{#2}}

\bibitem[{\citenamefont{Einstein et~al.}(1935)\citenamefont{Einstein, Podolsky,
  and Rosen}}]{epr}
\bibinfo{author}{\bibfnamefont{A.}~\bibnamefont{Einstein}},
  \bibinfo{author}{\bibfnamefont{B.}~\bibnamefont{Podolsky}}, \bibnamefont{and}
  \bibinfo{author}{\bibfnamefont{N.}~\bibnamefont{Rosen}},
  \bibinfo{journal}{Physical Review} \textbf{\bibinfo{volume}{47}},
  \bibinfo{pages}{777} (\bibinfo{year}{1935}).

\bibitem[{\citenamefont{Bohm}(1951)}]{bohm51}
\bibinfo{author}{\bibfnamefont{D.}~\bibnamefont{Bohm}},
  \emph{\bibinfo{title}{Quantum Theory}} (\bibinfo{publisher}{Prentice Hall,
  New York}, \bibinfo{year}{1951}).

\bibitem[{\citenamefont{Bell}(1964)}]{bell64}
\bibinfo{author}{\bibfnamefont{J.~S.} \bibnamefont{Bell}},
  \bibinfo{journal}{Physics} \textbf{\bibinfo{volume}{1}}, \bibinfo{pages}{195}
  (\bibinfo{year}{1964}), \bibinfo{note}{reprinted in \cite[p.~14]{bell87}}.

\bibitem[{\citenamefont{Bell}(1971)}]{bell71}
\bibinfo{author}{\bibfnamefont{J.~S.} \bibnamefont{Bell}}, in
  \emph{\bibinfo{booktitle}{Foundations of Quantum Mechanics}},
  \bibinfo{organization}{Proceedings of the International School of Physics
  `Enrico Fermi'} (\bibinfo{publisher}{New York, Academic},
  \bibinfo{year}{1971}), p. \bibinfo{pages}{171}, \bibinfo{note}{reprinted in
  \cite[p.~29]{bell87}.}

\bibitem[{\citenamefont{Bell}(1975)}]{bell76}
\bibinfo{author}{\bibfnamefont{J.~S.} \bibnamefont{Bell}},
  \emph{\bibinfo{title}{{The theory of local beables}}},
  \bibinfo{howpublished}{TH-2053-CERN} (\bibinfo{year}{1975}),
  \bibinfo{note}{presented at the Sixth GIFT Seminar, Jaca, 2--7 June 1975,
  reproduced in \emph{Epistemological Letters}, March 1976, and reprinted in
  \cite[p.~52]{bell87}}.

\bibitem[{\citenamefont{Hofer-Szab\'o et~al.}(1999)\citenamefont{Hofer-Szab\'o,
  R\'edei, and Szab\'o}}]{redei99}
\bibinfo{author}{\bibfnamefont{G.}~\bibnamefont{Hofer-Szab\'o}},
  \bibinfo{author}{\bibfnamefont{M.}~\bibnamefont{R\'edei}}, \bibnamefont{and}
  \bibinfo{author}{\bibfnamefont{L.~E.} \bibnamefont{Szab\'o}},
  \bibinfo{journal}{British Journal for the Philosophy of Science}
  \textbf{\bibinfo{volume}{50}}, \bibinfo{pages}{377} (\bibinfo{year}{1999}),
  \eprint{quant-ph/9805066}.

\bibitem[{\citenamefont{Reichenbach}(1956)}]{reichenbach56}
\bibinfo{author}{\bibfnamefont{H.}~\bibnamefont{Reichenbach}},
  \emph{\bibinfo{title}{The Direction of Time}} (\bibinfo{publisher}{University
  of California Press, Los Angeles}, \bibinfo{year}{1956}).

\bibitem[{\citenamefont{W\"uthrich}(2004)}]{wuethrich04}
\bibinfo{author}{\bibfnamefont{A.}~\bibnamefont{W\"uthrich}},
  \emph{\bibinfo{title}{Quantum Correlations and Common Causes}}
  (\bibinfo{publisher}{Bern Studies in the History and Philosophy of Science,},
  \bibinfo{year}{2004}).

\bibitem[{\citenamefont{Bell}(1987)}]{bell87}
\bibinfo{author}{\bibfnamefont{J.~S.} \bibnamefont{Bell}},
  \emph{\bibinfo{title}{Speakable and Unspeakable in Quantum Mechanics}}
  (\bibinfo{publisher}{Cambridge University Press, Cambridge},
  \bibinfo{year}{1987}).

\bibitem[{\citenamefont{Wigner}(1970)}]{wigner70}
\bibinfo{author}{\bibfnamefont{E.}~\bibnamefont{Wigner}},
  \bibinfo{journal}{American Journal of Physics} \textbf{\bibinfo{volume}{38}},
  \bibinfo{pages}{1005} (\bibinfo{year}{1970}).

\bibitem[{\citenamefont{van Fraassen}(1989)}]{vanfraassen89}
\bibinfo{author}{\bibfnamefont{B.~C.} \bibnamefont{van Fraassen}}, in
  \cite{cushing89}.

\bibitem[{\citenamefont{Clauser and Horne}(1974)}]{horne74}
\bibinfo{author}{\bibfnamefont{J.~F.} \bibnamefont{Clauser}} \bibnamefont{and}
  \bibinfo{author}{\bibfnamefont{M.~A.} \bibnamefont{Horne}},
  \bibinfo{journal}{Physical Review D} \textbf{\bibinfo{volume}{10}},
  \bibinfo{pages}{526} (\bibinfo{year}{1974}).

\bibitem[{\citenamefont{Butterfield}(1989)}]{butterfield89}
\bibinfo{author}{\bibfnamefont{J.}~\bibnamefont{Butterfield}}, in
  \cite{cushing89}.

\bibitem[{\citenamefont{Jarrett}(1984)}]{jarrett84}
\bibinfo{author}{\bibfnamefont{J.~P.} \bibnamefont{Jarrett}},
  \bibinfo{journal}{No\^us} \textbf{\bibinfo{volume}{18}}, \bibinfo{pages}{569}
  (\bibinfo{year}{1984}).

\bibitem[{\citenamefont{Jarrett}(1989)}]{jarrett89}
\bibinfo{author}{\bibfnamefont{J.~P.} \bibnamefont{Jarrett}}, in
  \cite{cushing89}, pp. \bibinfo{pages}{60--79}.

\bibitem[{\citenamefont{Shimony}(1993)}]{shimony93}
\bibinfo{author}{\bibfnamefont{A.}~\bibnamefont{Shimony}},
  \emph{\bibinfo{title}{Search for a naturalistic world view}},
  vol.~\bibinfo{volume}{2} (\bibinfo{publisher}{Cambridge University Press},
  \bibinfo{year}{1993}).

\bibitem[{\citenamefont{Hofer-Szab\'o et~al.}()\citenamefont{Hofer-Szab\'o,
  R\'edei, and Szab\'o}}]{redei03}
\bibinfo{author}{\bibfnamefont{G.}~\bibnamefont{Hofer-Szab\'o}},
  \bibinfo{author}{\bibfnamefont{M.}~\bibnamefont{R\'edei}}, \bibnamefont{and}
  \bibinfo{author}{\bibfnamefont{L.~E.} \bibnamefont{Szab\'o}},
  \emph{\bibinfo{title}{{Reichenbachian Common Cause Systems}}},
  \bibinfo{howpublished}{Forthcoming in \emph{International Journal of
  Theoretical Physics}},
  \urlprefix\url{http://philsci-archive.pitt.edu/archive/00001246/}.

\bibitem[{\citenamefont{Placek}(2000)}]{placek00b}
\bibinfo{author}{\bibfnamefont{T.}~\bibnamefont{Placek}},
  \emph{\bibinfo{title}{Is Nature determinstic?}}
  (\bibinfo{publisher}{Jagiellonian University Press, Krak\'ow},
  \bibinfo{year}{2000}), \bibinfo{edition}{1st} ed.

\bibitem[{\citenamefont{Szab\'o}(1998)}]{Szabo:1998pb}
\bibinfo{author}{\bibfnamefont{L.~E.} \bibnamefont{Szab\'o}}
  (\bibinfo{year}{1998}), \eprint{quant-ph/9806074}.

\bibitem[{\citenamefont{Suppes and Zanotti}(1976)}]{suppes76}
\bibinfo{author}{\bibfnamefont{P.}~\bibnamefont{Suppes}} \bibnamefont{and}
  \bibinfo{author}{\bibfnamefont{M.}~\bibnamefont{Zanotti}}, in
  \emph{\bibinfo{booktitle}{Logic and Probability in Quantum Mechanics}},
  edited by \bibinfo{editor}{\bibfnamefont{P.}~\bibnamefont{Suppes}}
  (\bibinfo{publisher}{Dordrecht: Reidel}, \bibinfo{year}{1976}), pp.
  \bibinfo{pages}{445--455}.

\bibitem[{\citenamefont{Fine}(1982)}]{fine82}
\bibinfo{author}{\bibfnamefont{A.}~\bibnamefont{Fine}},
  \bibinfo{journal}{Synthese} \textbf{\bibinfo{volume}{50}},
  \bibinfo{pages}{279} (\bibinfo{year}{1982}).

\bibitem[{\citenamefont{Szab\'o}(2000)}]{szabo00}
\bibinfo{author}{\bibfnamefont{L.}~\bibnamefont{Szab\'o}}
  (\bibinfo{year}{2000}), \eprint{quant-ph/0002030}.

\bibitem[{\citenamefont{Szab\'o and Fine}(2002)}]{szabo02}
\bibinfo{author}{\bibfnamefont{L.}~\bibnamefont{Szab\'o}} \bibnamefont{and}
  \bibinfo{author}{\bibfnamefont{A.}~\bibnamefont{Fine}}
  (\bibinfo{year}{2002}),
  \urlprefix\url{http://philsci-archive.pitt.edu/archive/00000642/}.

\bibitem[{\citenamefont{Gra{\ss}hoff and May}(2001)}]{grasshoff01}
\bibinfo{author}{\bibfnamefont{G.}~\bibnamefont{Gra{\ss}hoff}}
  \bibnamefont{and} \bibinfo{author}{\bibfnamefont{M.}~\bibnamefont{May}}, in
  \emph{\bibinfo{booktitle}{Current Issues in Causation}}, edited by
  \bibinfo{editor}{\bibfnamefont{W.}~\bibnamefont{Spohn}},
  \bibinfo{editor}{\bibfnamefont{M.}~\bibnamefont{Ledwig}}, \bibnamefont{and}
  \bibinfo{editor}{\bibfnamefont{M.}~\bibnamefont{Esfeld}}
  (\bibinfo{publisher}{Paderborn: Mentis}, \bibinfo{year}{2001}), pp.
  \bibinfo{pages}{85--114}.

\bibitem[{\citenamefont{Gra{\ss}hoff and
  Baumgartner}(forthcoming)}]{gg_baumgartner}
\bibinfo{author}{\bibfnamefont{G.}~\bibnamefont{Gra{\ss}hoff}}
  \bibnamefont{and}
  \bibinfo{author}{\bibfnamefont{M.}~\bibnamefont{Baumgartner}},
  \emph{\bibinfo{title}{Kausalit\"at und kausales Schliessen}}
  (\bibinfo{publisher}{Bern Studies in the History and Philosophy of Science},
  \bibinfo{year}{forthcoming}).

\bibitem[{\citenamefont{Aspect et~al.}(1982)\citenamefont{Aspect, Dalibard, and
  Roger}}]{aspect82}
\bibinfo{author}{\bibfnamefont{A.}~\bibnamefont{Aspect}},
  \bibinfo{author}{\bibfnamefont{J.}~\bibnamefont{Dalibard}}, \bibnamefont{and}
  \bibinfo{author}{\bibfnamefont{G.}~\bibnamefont{Roger}},
  \bibinfo{journal}{Physical Review Letters} \textbf{\bibinfo{volume}{49}},
  \bibinfo{pages}{1804} (\bibinfo{year}{1982}).

\bibitem[{\citenamefont{Spirtes et~al.}(1993)\citenamefont{Spirtes, Glymour,
  and Scheines}}]{spirtes93}
\bibinfo{author}{\bibfnamefont{P.}~\bibnamefont{Spirtes}},
  \bibinfo{author}{\bibfnamefont{C.}~\bibnamefont{Glymour}}, \bibnamefont{and}
  \bibinfo{author}{\bibfnamefont{R.}~\bibnamefont{Scheines}},
  \emph{\bibinfo{title}{Causation, Prediction, and Search}}
  (\bibinfo{publisher}{Springer Verlag}, \bibinfo{year}{1993}).

\bibitem[{\citenamefont{Simpson}(1951)}]{simpson51}
\bibinfo{author}{\bibfnamefont{E.~H.} \bibnamefont{Simpson}},
  \bibinfo{journal}{Journal of the Royal Statistical Society, Ser. B.}
  \textbf{\bibinfo{volume}{13}}, \bibinfo{pages}{238} (\bibinfo{year}{1951}).

\bibitem[{\citenamefont{Cartwright}(1979)}]{cartwright79}
\bibinfo{author}{\bibfnamefont{N.}~\bibnamefont{Cartwright}},
  \bibinfo{journal}{No\^us} \textbf{\bibinfo{volume}{13}}, \bibinfo{pages}{419}
  (\bibinfo{year}{1979}).

\bibitem[{\citenamefont{Cushing and McMullin}(1989)}]{cushing89}
\bibinfo{editor}{\bibfnamefont{J.~T.} \bibnamefont{Cushing}} \bibnamefont{and}
  \bibinfo{editor}{\bibfnamefont{E.}~\bibnamefont{McMullin}}, eds.,
  \emph{\bibinfo{title}{Philosophical Consequences of Quantum Theory.
  Reflections on Bell's Theorem}} (\bibinfo{publisher}{University of Notre Dame
  Press, Notre Dame}, \bibinfo{year}{1989}).

\end{thebibliography}

\end{document}